\documentclass{article}
\usepackage{ace,amsmath,graphicx,url,times}
\usepackage{psfrag}

\title{Joint Estimation of Reverberation Time and Direct-to-Reverberation Ratio from Speech using Auditory-Inspired Features}

\name{Feifei Xiong$^{1,3}$, Stefan Goetze$^{1,3}$, Bernd T. Meyer$^{2,3}$}
\address
{
$^{1}$Fraunhofer Institute for Digital Media Technology IDMT,\\
Project Group Hearing, Speech and Audio Technology (HSA), Oldenburg, Germany\\
$^{2}$Medizinische Physik, Carl von Ossietzky University Oldenburg, Germany\\
$^{3}$Cluster of Excellence Hearing4all, Carl von Ossietzky University Oldenburg, Germany\\
\normalsize{\texttt{\{feifei.xiong,s.goetze\}@idmt.fraunhofer.de,}\, \texttt{bernd.meyer@uni-oldenburg.de}}
}

\begin{document}

\ninept
\maketitle

\begin{sloppy}

\begin{abstract}
Blind estimation of acoustic room parameters such as the reverberation time~$T_\mathrm{60}$ and the direct-to-reverberation ratio~($\mathrm{DRR}$) 
is still a challenging task, especially in case of blind estimation from reverberant speech signals.
In this work, a novel approach is proposed for joint estimation of $T_\mathrm{60}$ and $\mathrm{DRR}$ from wideband speech in noisy conditions.
2D Gabor filters arranged in a filterbank are exploited for extracting features, 
which are then used as input to a multi-layer perceptron~(MLP).
The MLP output neurons correspond to specific pairs of $(T_\mathrm{60}, \mathrm{DRR})$ estimates; 
the output is integrated over time, and a simple decision rule results in our estimate.
The approach is applied to single-microphone fullband speech signals provided by the 
Acoustic Characterization of Environments~(ACE) Challenge.
Our approach outperforms the baseline systems with median errors of close-to-zero and -1.5~dB for the
$T_\mathrm{60}$ and $\mathrm{DRR}$ estimates, respectively, 
while the calculation of estimates is 5.8 times faster compared to the baseline.
\end{abstract}

\begin{keywords}
Reverberation time, direct-to-reverberation ratio, 2D Gabor features, multi-layer perceptron, ACE Challenge
\end{keywords}

\section{Introduction}
\label{sec:intro}
The acoustic characteristics of a room have been shown to be important 
to predict the speech quality and intelligibility, 
which is relevant to speech enhancement~\cite{NG10} as well as for automatic speech recognition~(ASR)~\cite{Seh09}.
The reverberation time~$T_\mathrm{60}$ and the direct-to-reverberation ratio~($\mathrm{DRR}$)
are two important acoustic parameters.
Traditionally, $T_\mathrm{60}$ and $\mathrm{DRR}$ can be obtained from a measured room impuls response~(RIR)~\cite{Kut00}.
However, it is not practical or not even possible to measure the corresponding RIRs in most applications. 
Consequently, the demand of blind $T_\mathrm{60}$ and $\mathrm{DRR}$ estimation directly from speech and audio signals is increasing.

A number of approaches for blind estimation have been proposed earlier:
Based on the spectral decay distribution of the reverberant signal, 
$T_\mathrm{60}$ is determined in~\cite{WHN08} by estimating the decay rate in each frequency band.
A noise-robust version is presented in~\cite{EGN13}.
In~\cite{RJWO+03} a blind $T_\mathrm{60}$ estimation is achieved by a statistical model of the sound decay characteristics of reverberant speech.
Inspired by this, \cite{LYJV10} uses a pre-selection mechanism to detect plausible decays and 
a subsequent application of a maximum-likelihood criterion to estimate $T_\mathrm{60}$ with a low computational complexity.
Alternatively, motivated by the progress that has been achieved using artificial neural networks in machine learning tasks,
\cite{CLD01}~proposed a method to estimate $T_\mathrm{60}$ blindly from reverberant speech using trained neural networks, 
for which short-term root-mean square values of speech signals were used as the network input.
The approach in~\cite{CLD01} was also extended to estimate various acoustic room parameters in~\cite{KCLZ+08} using the low frequency envelope spectrum.
Our work~\cite{XGM13} proposed a multi-layer perceptron using spectro-temporal modulation features to estimate $T_\mathrm{60}$.
A comparison of energies at high and low modulation frequencies, the so-called speech-to-reverberation modulation energy ratio~(SRMR), 
which is highly correlated to $T_\mathrm{60}$ and $\mathrm{DRR}$, is evaluated in~\cite{FC10}.

The approaches mentioned so far use a single audio channel for obtaining the estimate,
however, the majority of blind off-the-shelf $\mathrm{DRR}$ estimators rely on multi-channel data. 
An approach to estimate $\mathrm{DRR}$ based on a binaural input signal
from which the direct component is eliminated by an equalization-cancellation operation was proposed in~\cite{LC08}.
Another method using an octagonal microphone array has been presented in~\cite{HNSF+11}, 
where a spatial coherence matrix for the mixture of a direct and diffuse sound field was employed to estimate $\mathrm{DRR}$ using a least-squares criterion.
In~\cite{Kus11}, an analytical expression was derived for the relationship between the $\mathrm{DRR}$ and the binaural magnitude-squared coherence function.
A null-steering beamformer is employed in~\cite{EMNS15} to estimate the $\mathrm{DRR}$ with a two-element microphone array.

Motivated by the fact that the amount of perceived reverberation depends on both $T_\mathrm{60}$ and $\mathrm{DRR}$,
we propose a novel approach to simultaneously and blindly estimate these parameters.
In our previous work~\cite{XGM13,XGM14}, we found spectro-temporal modulation features obtained by a 2D Gabor filterbank
to be strongly and non-linearly correlated with reverberation parameters.
We refer to these features as \emph{auditory} Gabor features, 
since the filters used for extraction resemble the spectro-temporal receptive fields 
in the auditory cortex of mammals~\cite{QSE03}, i.e., 
it is likely that our auditory system is explicitly tuned to such patterns. 
The Gabor features are used as input to an artificial neural network, i.e.~a multi-layer perceptron~(MLP),
which is trained for blind estimation of the parameters pair $(T_\mathrm{60}, \mathrm{DRR})$.
The evaluation of performance focuses on the Acoustic Characterization of Environments~(ACE) Challenge~\cite{EGMN15} 
evaluation test set in fullband mode with a single microphone.

The remainder of this paper is organized as follows:
Section~\ref{sec:method} introduces the blind $(T_\mathrm{60}, \mathrm{DRR})$ estimator
based on the 2D Gabor features and an MLP classifier.
The detailed experimental procedure is described in Section~\ref{sec:expm}
according to the ACE Challenge regulations.
The results and discussion are presented in Section~\ref{sec:result} 
for the proposed $(T_\mathrm{60}, \mathrm{DRR})$ estimator with the ACE evaluation test set,
and Section~\ref{sec:conclu} concludes the paper.

\section{Blind $(T_\mathrm{60}, \mathrm{DRR})$ Estimator}
\label{sec:method}
An overview of the estimation process is presented in Figure~\ref{fig:mlp}: 
In a first step, reverberant signals are converted to spectro-temporal Gabor filterbank features~\cite{MRSM11,SMK12}
to capture information relevant for room parameters estimation. 
An MLP is trained to map the input pattern to pairs of parameters $(T_\mathrm{60}, \mathrm{DRR})$,
where the label information is according to $(T_\mathrm{60}, \mathrm{DRR})$ from the available RIRs.
Since the MLP generates one estimate per time step, we obtain an \emph{utterance}-based estimate 
by simple temporal averaging and subsequent selection of the output neuron with the highest average activation
(\emph{winner-takes-all}), as shown in Figure~\ref{fig:MLP_label} for instance.
The noisy reverberant speech signal $y[k]$ is constructed from clean (anechoic) speech $s[k]$ 
convolved with measured RIRs $h[k]$ and an additive noise $n[k]$, denoted as $y[k] = s[k] \ast h[k] + n[k]$ with time index $k$.
\begin{figure}[h!]
  \centering
  \psfrag{y[k]}  			 			[c][c]{\scriptsize $y[k]$}
	\psfrag{n[k]}  			 			[c][c]{\scriptsize $n[k]$}
	\psfrag{s[k]}  			 			[c][c]{\scriptsize $s[k]$}
	\psfrag{h[k]}  			 			[c][c]{\scriptsize $h[k]$}
	\psfrag{available}   			[c][c]{\scriptsize available}
  \psfrag{RIRs}           	[c][c]{\scriptsize RIRs}  
	\psfrag{clean}     			  [c][c]{\scriptsize anechoic} 
	\psfrag{speech}     			[c][c]{\scriptsize speech} 
	\psfrag{noise}            [c][c]{\scriptsize noise} 
	\psfrag{GBFB}  	          [c][c]{\scriptsize Gabor features}
	\psfrag{feature}          [c][c]{\scriptsize feature}
	\psfrag{extraction}       [c][c]{\scriptsize extraction}
	\psfrag{MLP}     				  [c][c]{\scriptsize MLP}
	\psfrag{label}   				  [c][c]{\scriptsize Label}
	\psfrag{temporal} 			  [c][c]{\scriptsize temporal}
	\psfrag{average} 				  [c][c]{\scriptsize average}
	\psfrag{T60, DRR} 			  [c][c]{\tiny $(T_\mathrm{60}, \mathrm{DRR})$}
  \centerline{\includegraphics[width=\columnwidth]{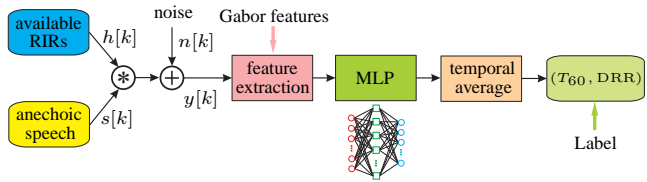}}
  \caption{Overview of the MLP setup for $(T_\mathrm{60}, \mathrm{DRR})$ estimation.}
  \label{fig:mlp}
\end{figure}
\begin{figure}[h] 
\centering
 \psfrag{100}          [r][r]{\tiny 100}
 \psfrag{80}           [r][r]{\tiny 80}
 \psfrag{60}           [r][r]{\tiny 60}
 \psfrag{40}           [r][r]{\tiny 40}
 \psfrag{20}           [r][r]{\tiny 20}
 \psfrag{0.1}          [c][c]{\tiny 0.1}
 \psfrag{0.3}          [c][c]{\tiny 0.3}
 \psfrag{0}            [c][c]{\tiny 0}
 \psfrag{0.8}          [c][c]{\tiny 0.8}
 \psfrag{0.6}          [c][c]{\tiny 0.6}
 \psfrag{0.4}          [c][c]{\tiny 0.4}
 \psfrag{0.2}          [c][c]{\tiny 0.2}
 \psfrag{time frames}     							[c][c]{\tiny frames in time slot}
 \psfrag{mean probability}  						[c][c]{\tiny mean probability}
 \psfrag{class}     								    [c][c]{\tiny Labels}
 \psfrag{(a) MLP output}     						[c][c]{\tiny (a) MLP output}
 \psfrag{(b) decision}     						  [c][c]{\tiny (b) mean value across frames}
 \resizebox{1\linewidth}{!}{\includegraphics[width=0.8\linewidth]{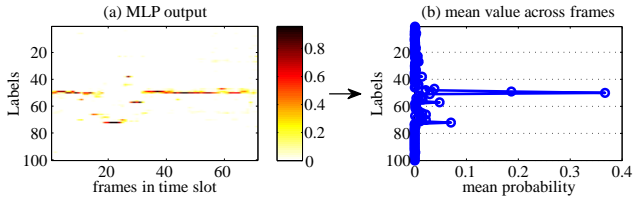}}
 \caption{Visualization of the decision rule applied to convert the frame-wise MLP-output
 to a $(T_\mathrm{60}, \mathrm{DRR})$ class for each utterance.}
\label{fig:MLP_label}
\end{figure}

Gabor features are generated by 2D Gabor filters ${G_b}$ applied to filter log-mel-spectrograms.
The filters ${G_b}$ are localized spectro-temporal patterns that are with a high sensitivity towards amplitude modulations,
as defined by
\begin{eqnarray}
{G_b}[m, \ell]  &=& {S}_\mathrm{carr} [m, \ell] \cdot {H}_{\mathrm{env}} [m, \ell]  \, , \label{eqn:gabor}\\
{S}_\mathrm{carr} [m, \ell]  &=& \exp{(i \omega_m(m-m_0) + i \omega_\ell(\ell-\ell_0))} \label{eqn:gabor_carr}  \, , \\
{H}_{\mathrm{env}} [m,\ell]  &=& 0.5-0.5 \cdot \cos\left( \frac{2\pi(m-m_0)}{W_m+1} \right)  \nonumber \\
                & &   \hspace{37.6pt}		 \cdot \cos\left( \frac{2\pi(\ell-\ell_0)}{W_\ell+1} \right)  \, , \label{eqn:gabor_env}
\end{eqnarray}
with $m$ and $\ell$ denoting the (mel-)spectral and temporal frame indices, and $W_m,\,W_\ell$ 
the Hann-envelope ${H}_{\mathrm{env}}$ window lengths with the center indices $m_0,\, \ell_0$, respectively.
The periodicity of the sinusoidal-carrier function ${S}_\mathrm{carr}$ is defined
by the radian frequencies $\omega_m,\, \omega_\ell$, which allow the Gabor filters to be tuned
to particular directions of spectro-temporal modulation.
The purely diagonal Gabor filters as shown in Figure~\ref{fig:gbfb_2d}, 
were found to result in the maximal sensitivity to the reverberation effect~\cite{XGM13}
and thus, are used here to construct the Gabor features for the $(T_\mathrm{60}, \mathrm{DRR})$ estimator.
Each log-mel-spectrogram is filtered with these 48 filters in the filterbank
that cover temporal modulations from 2.4 to 25~Hz~and spectral modulations from -0.25 to 0.25~cycles/channel, respectively.
\begin{figure}[t] 
  \centering
	\psfrag{25.0}         [c][c]{\scriptsize 25.0}
  \psfrag{15.7}         [c][c]{\scriptsize 15.7}
  \psfrag{9.9}          [c][c]{\scriptsize 9.9}
  \psfrag{6.2}          [c][c]{\scriptsize 6.2}
  \psfrag{3.9}          [c][c]{\scriptsize 3.9}
  \psfrag{2.4}          [c][c]{\scriptsize 2.4}
  \psfrag{0.0}          [c][c]{\scriptsize 0.0}
  \psfrag{0.25}         [l][l]{\scriptsize 0.25}
  \psfrag{0.12}         [l][l]{\scriptsize 0.12}
  \psfrag{0.06}         [l][l]{\scriptsize 0.06}
  \psfrag{0.03}         [l][l]{\scriptsize 0.03}
  \psfrag{0}            [l][l]{\scriptsize 0.0}
  \psfrag{-0.25}        [l][l]{\scriptsize -0.25}
  \psfrag{-0.12}        [l][l]{\scriptsize -0.12}
  \psfrag{-0.06}        [l][l]{\scriptsize -0.06}
  \psfrag{-0.03}        [l][l]{\scriptsize -0.03}
	\psfrag{2D GBFB}  					        			[c][c]{\scriptsize Diagonal Gabor filters}
  \psfrag{temporal mod. / Hz}          			[c][c]{\scriptsize Temporal modulation frequency [Hz]}
  \psfrag{spectral mod. / cycl./chan.}      [c][c]{\scriptsize Spectral modulation frequency [cycl./chan.]} 
  \centerline{\includegraphics[width=\columnwidth]{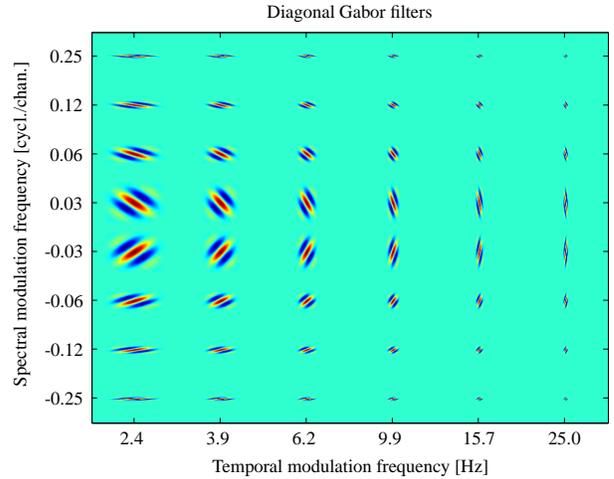}}
  \caption{Real component of the 2D Gabor filterbank set with only diagonal Gabor filters.}
  \label{fig:gbfb_2d}
\end{figure}

\section{Experimental Setup}
\label{sec:expm}
\subsection{ACE Challenge}
\label{ssec:ace}
The ACE Challenge provides a development~(Dev)~dataset for algorithm fine-tuning
and an evaluation~(Eval)~dataset for the final algorithm test.
The task is aiming at blindly estimating two acoustic parameters, i.e.~$T_\mathrm{60}$ and $\mathrm{DRR}$, from noisy and reverberant speech.
Two different modes
i.e.~fullband and subband (1/3-octave ISO~\cite{ANSI93} since $T_\mathrm{60}$ and $\mathrm{DRR}$ are both frequency dependent parameters),
and six microphone configurations, i.e.~a single microphone~(Single) and microphone arrays with 
two~(Laptop), three~(Mobile), five~(Cruciform), eight~(Linear), and thirty-two~(Spherical) microphones, were introduced.
The dataset was generated using anechoic speech convolved with RIRs measured from real rooms with additive noise recorded under the same conditions.
Also, three types of noise signals, i.e.~ambient, babble and fan noises, were added to generate the noisy reverberant dataset.
For Dev dataset, the signal-to-noise ratios~(SNRs) were chosen to be 0, 10 and 20~dB,
while for Eval, the SNRs were -1, 12 and 18~dB.
The Dev dataset is approximately 120~h length from all multi-microphone scenarios.
Each test set from Eval contains 4500 utterances categorized by these 3~noise types and 3~SNRs.
For this paper, we focus on the tasks in the fullband mode of $T_\mathrm{60}$ and $\mathrm{DRR}$ estimation in the single microphone scenarios.
Our approach is also applicable to multi-microphone scenarios by selecting any channel of the speech data.

The ground truth values of $T_\mathrm{60}$ and $\mathrm{DRR}$ were provided by the ACE Challenge.
The ground truth $T_\mathrm{60}$ is based on the energy decay curve computed from the RIRs using the Schroeder integral~\cite{Sch65},
to which the method proposed in~\cite{KAMP+02} is used to estimate $T_\mathrm{60}$. 
This method is shown to be more reliable under all conditions than the standard method according to ISO3382~\cite{ISO09}.
The ground truth $\mathrm{DRR}$ is estimated using the method of~\cite{MSGE12},
where the direct path is determined by the $\pm$8~ms around the maximum found using an equalization filter~\cite{EGMN15}.

\subsection{Platform}
\label{ssec:plat}
The MLP shown in Figure~\ref{fig:mlp} was implemented with the open-source Kaldi ASR toolkit~\cite{PGBB+11}
compiled with a Tesla K20c NVIDIA GPU with 5~GB memory size.
It had 3~layers:
The number of neurons in the input layer is 600, i.e.~dimension of the 2D diagonal Gabor features 
(cf.~Figure~\ref{fig:gbfb_2d}) calculated in Matlab.
The temporal context considered by the MLP is limited to 1~frame, i.e.~no splicing is applied.
The number of hidden units is a free parameter that was optimized given the amount of training data and set to 8192 units, 
and the number of output neurons corresponds to the amount of $(T_\mathrm{60}, \mathrm{DRR})$ pairs,
i.e.~100 as defined in the following (also cf.~Figure~\ref{fig:MLP_label}).

\subsection{Speech Database}
\label{ssec:speech}
ACE database was recorded by different individuals who were reading different text materials in English.
Here, we applied TIMIT corpus~\cite{GLFF+93} to generate the training data for MLP,
since TIMIT contains recordings of phonetically-balanced prompted English speech
and a total of 6300~sentences (approximately 5.4~h).
To avoid a strong mismatch between training and test data 
(which is likely to hurt MLP classification performance) 
we added the ACE Dev dataset to the training data.
In order to match the amount of the Dev dataset (approximately 120~h), 
thereby balancing the two sets,
TIMIT utterances were convolved with the collected RIRs circularly,
which resulted in approximately 117~h TIMIT training data.
The sampling rate of all signals is 16~kHz.

\subsection{RIR Database}
\label{ssec:rir}
\begin{figure}[b!]
  \centering
	\psfrag{-6}         [c][c]{\scriptsize -6}
	\psfrag{-5}         [c][c]{\scriptsize -5}
	\psfrag{-4}         [c][c]{\scriptsize -4}
	\psfrag{-3}         [c][c]{\scriptsize -3}
	\psfrag{-2}         [c][c]{\scriptsize -2}
	\psfrag{-1}         [c][c]{\scriptsize -1}
	\psfrag{0}          [c][c]{\scriptsize 0}
	\psfrag{1}          [c][c]{\scriptsize 1}
	\psfrag{2}          [c][c]{\scriptsize 2}
	\psfrag{3}          [c][c]{\scriptsize 3}
	\psfrag{4}          [c][c]{\scriptsize 4}
	\psfrag{5}          [c][c]{\scriptsize 5}
	\psfrag{6}          [c][c]{\scriptsize 6}
	\psfrag{7}          [c][c]{\scriptsize 7}
	\psfrag{8}          [c][c]{\scriptsize 8}
	\psfrag{9}          [c][c]{\scriptsize 9}
	\psfrag{10}         [c][c]{\scriptsize 10}
	\psfrag{11}         [c][c]{\scriptsize 11}
	\psfrag{12}         [c][c]{\scriptsize 12}
	\psfrag{13}         [c][c]{\scriptsize 13}
	\psfrag{14}         [c][c]{\scriptsize 14}
	\psfrag{15}         [c][c]{\scriptsize 15}
  \psfrag{1350}         [r][r]{\scriptsize 1350}
	\psfrag{1250}         [r][r]{\scriptsize 1250}
	\psfrag{1150}         [r][r]{\scriptsize 1150}
	\psfrag{850}          [r][r]{\scriptsize 850}
	\psfrag{750}          [r][r]{\scriptsize 750}
	\psfrag{650}          [r][r]{\scriptsize 650}
	\psfrag{550}          [r][r]{\scriptsize 550}
	\psfrag{450}          [r][r]{\scriptsize 450}
	\psfrag{350}          [r][r]{\scriptsize 350}
	\psfrag{250}          [r][r]{\scriptsize 250}
	\psfrag{150}          [r][r]{\scriptsize 150}
	\psfrag{T}          [c][c]{\scriptsize $T_\mathrm{60}$ / ms}
	\psfrag{R}          [c][c]{\scriptsize $\mathrm{DRR}$ / dB}
	\psfrag{D}          [c][c]{\scriptsize $(T_\mathrm{60}, \mathrm{DRR})$ Distribution}
	\psfrag{collected}          [l][l]{\tiny Collected RIRs}
	\psfrag{dev}   			        [l][l]{\tiny ACE Dev}
	\psfrag{eval}      			    [l][l]{\tiny ACE Eval}
  \centerline{\includegraphics[width=\columnwidth]{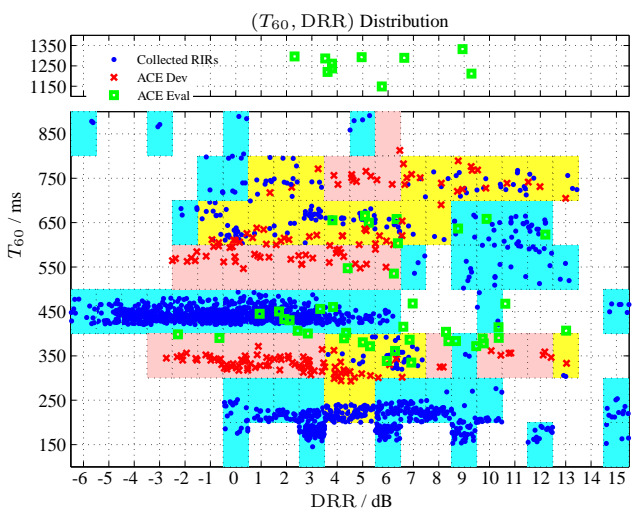}}
  \caption{Distribution of the open-source RIR databases with respect to $(T_\mathrm{60}, \mathrm{DRR})$, 
	         as well as the ground truth values of the ACE Dev and Eval datasets.}
  \label{fig:rir_distru}
\end{figure}
To cover a wide range of RIRs that occur in real life scenarios, 
we use several open-source RIR databases such as 
MARDY~\cite{WGHM+06}, AIR database~\cite{JSV09}, REVERB Challenge~\cite{Kin13} and SMARD~\cite{NJJC14}.
Further, we also recorded several RIRs in two regular office rooms in our group.
Figure~\ref{fig:rir_distru} shows the distribution of $(T_\mathrm{60}, \mathrm{DRR})$ values from the collected RIRs,
as well as the ACE Dev and Eval datasets.
$(T_\mathrm{60}, \mathrm{DRR})$ ground truth values of the collected RIRs 
were calculated based on the methods described in Section~\ref{ssec:ace}.
Due to the lack of the corresponding equalization filters for the source,
the absolute peak position is considered as the maximum to determine the direct path for the $\mathrm{DRR}$ calculations.

An MLP has a limited number of output neurons, 
which limits the resolution for the target estimate. 
We chose a resolution based on the distribution of training RIRs, 
with the aim of obtaining a sufficient number of $(T_\mathrm{60}, \mathrm{DRR})$ observations for each pair, 
which is 100~ms for $T_\mathrm{60}$ and 1~dB for $\mathrm{DRR}$ (cf.~Figure~\ref{fig:rir_distru} where one bounding box represents one class).
The boundaries of $T_\mathrm{60}$ are 100~ms and 900~ms, with $\mathrm{DRR}$ ranging from -6~dB to 15~dB. 
With these boundaries and the chosen resolution, 
76~classes are obtained for the collected RIRs (light blue boxes), and 51~classes are obtained from the ACE Dev dataset (light red boxes). 
These classes are partially overlapping (light yellow boxes) and result in a total of 100~classes.

\subsection{Noise Signals}
\label{ssec:noise}
The ACE noise signals were recorded in the same acoustic conditions as the RIR measurement,~i.e., the noise captured by the microphone is reverberated. Hence, the noise signals combined with our extended RIRs should be reverberated as well. Since the original noise signals were not available in the context of the challenge, we created noise signals with similar characteristics as the original ambient, babble and fan noise. 
\begin{itemize}
	\item Ambient noise was created by mixing recorded car noise and pink noise to obtain a colored noise with high energy in the low frequencies
	(as the original ambient noise).
	\item To create babble noise, we mixed clean speech signals (two male, two female speakers) from the WSJCAM0~\cite{RFPFR95} corpus. 
	\item A fan noise was recorded in an almost anechoic chamber to obtain the last noise type.
\end{itemize}
Subsequently, the noise signals were added to the anechoic speech at SNRs of 0, 10 and 20~dB 
(mimicking the procedure for the ACE Dev dataset), which were then convolved with the collected RIRs.

\section{Results}
\label{sec:result}
The estimation error is used for analysis 
and is defined as the difference between the estimated value and the ground truth value,
i.e.~$E_{T_\mathrm{60}} = \widehat{T_\mathrm{60}} - {T_\mathrm{60}}$ in~s for $T_\mathrm{60}$
and $E_{\mathrm{DRR}} = \widehat{\mathrm{DRR}} - {\mathrm{DRR}}$ in~dB for $\mathrm{DRR}$.
For comparison,
the methods proposed in~\cite{LYJV10} and in~\cite{FC10} 
are employed as baseline to blindly estimate $T_\mathrm{60}$ and $\mathrm{DRR}$, respectively.
Note that the blind $\mathrm{DRR}$ estimator in~\cite{FC10} requires a mapping function
between the overall SRMR from 5th to 8th channel and the $\mathrm{DRR}$ (both expressed in~dB),
which is obtained by the ACE Dev Single dataset.
\begin{figure}[b!]
  \centering
	\psfrag{A}						      [c][c]{\tiny Ambient}
	\psfrag{B}						      [c][c]{\tiny Babble}
	\psfrag{F}						      [c][c]{\tiny Fan}
	\psfrag{T}						      [c][c]{\tiny $E_{T_\mathrm{60}}$ / s}
	\psfrag{D}						      [c][c]{\tiny $E_{\mathrm{DRR}}$ / dB}
	\psfrag{S}						      [c][c]{\tiny SNR / dB}
	\psfrag{baseline}		        [l][l]{\tiny baseline}
	\psfrag{mlp}	    	        [l][l]{\tiny proposed}
  \centerline{\includegraphics[width=\columnwidth]{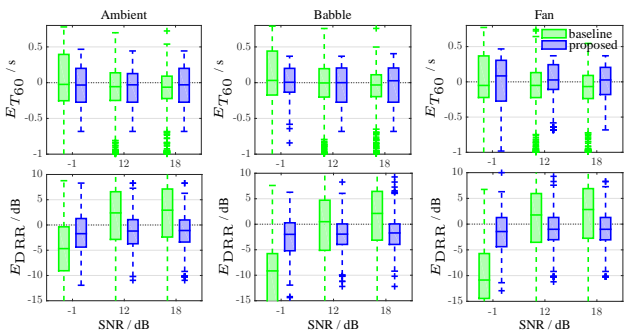}}
  \caption{Performance of $(T_\mathrm{60}, \mathrm{DRR})$ estimation for ACE Eval Single dataset.
					On each box, the central mark is the median, the edges are the 25th and 75th percentiles,
					the whiskers show extreme values and the outliers are plotted individually.}
  \label{fig:t60_single_comp}
\end{figure}
\begin{figure*}[t!]
  \centering
	\psfrag{laptop}		          [l][l]{\scriptsize Laptop}
	\psfrag{mobile}		          [l][l]{\scriptsize Mobile}
	\psfrag{cruciform}		      [l][l]{\scriptsize Cruciform}
	\psfrag{linear}				      [l][l]{\scriptsize Linear}
	\psfrag{spherical}		      [l][l]{\scriptsize Spherical}
	\psfrag{A}						      [c][c]{\scriptsize Ambient}
	\psfrag{B}						      [c][c]{\scriptsize Babble}
	\psfrag{F}						      [c][c]{\scriptsize Fan}
	\psfrag{T}						      [c][c]{\scriptsize $E_{T_\mathrm{60}}$ / s}
	\psfrag{D}						      [c][c]{\scriptsize $E_{\mathrm{DRR}}$ / dB}
	\psfrag{S}						      [c][c]{\scriptsize SNR / dB}
  \centerline{\includegraphics[width=2.1\columnwidth]{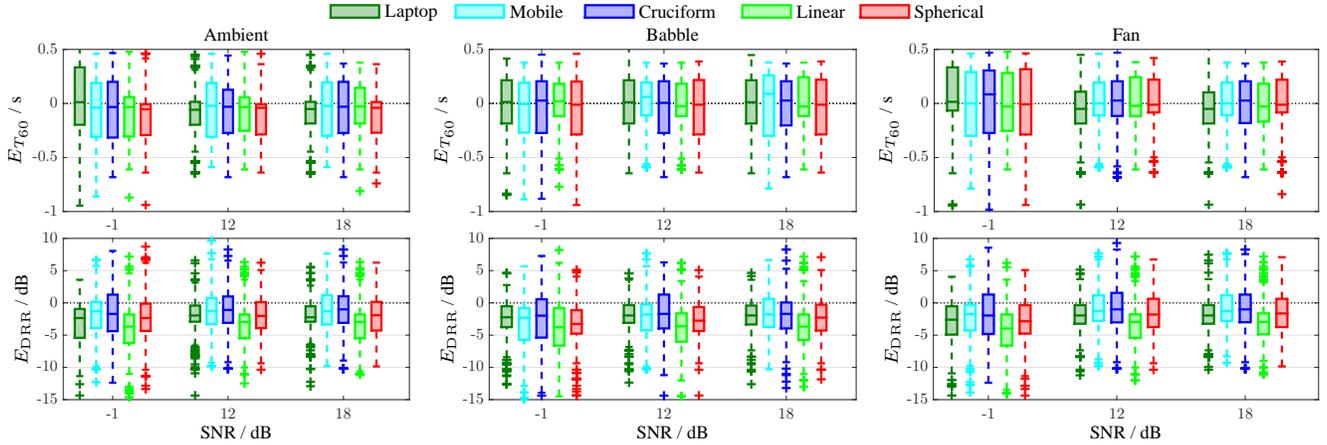}}
  \caption{Performance of $(T_\mathrm{60}, \mathrm{DRR})$ estimation for ACE Eval multi-microphone scenarios,
					 i.e.~Laptop, Mobile, Cruciform, Linear and Spherical (cf.~Section~\ref{ssec:ace}) with the first channel~(\emph{ch1}) data.}
  \label{fig:multi_ch1}
	\vspace{-1.0em}
\end{figure*}

As seen in Figure~\ref{fig:t60_single_comp}, in general, the proposed method outperforms the baseline approaches.
For $T_\mathrm{60}$, the baseline method works better in slightly noisy environments with an SNR of 18~dB,
while the performances degrade with lower SNRs.
The proposed method 
has a higher robustness with respect to additive noise, presumably because the statistical model is trained on noisy reverberant speech with various SNRs.
The median values of $E_{T_\mathrm{60}}$ are close to 0~ms for all conditions (3~noise types and 3~SNRs), 
and the upper and lower percentiles are within $\pm$250~ms,
which indicates that the proposed method is capable of providing accurate blind $T_\mathrm{60}$ estimation.
In addition, far less outliers are obtained compared to the baseline method.
The same trend can be observed for $E_{\mathrm{DRR}}$,
for which the baseline produces large errors for both median and percentiles,
particularly in the low SNR situations.

The $\mathrm{DRR}$ is underestimated by approximately -1.5~dB.
This could be explained by the limited resolution of estimates 
(100~ms for $E_{T_\mathrm{60}}$, 1~dB for $E_{\mathrm{DRR}}$)
and the mismatch of data range for training data on the one hand, 
and for Eval dataset on the other:
As shown in Figure~\ref{fig:rir_distru}, $T_\mathrm{60}$ values from 1100~ms to 1400~ms 
are not covered by the training data at all. 
A detailed post analysis showed that underestimates of $T_\mathrm{60}$ that 
arise from this mismatch go along with underestimates of $\mathrm{DRR}$; 
for instance, a test sample with ground truth of (1293~ms,~4.96~dB) was estimated to be (750~ms,~1~dB).
It appears that the underestimated reverberation effect caused by an underestimate of $T_\mathrm{60}$ 
is somehow compensated by the corresponding underestimate of $\mathrm{DRR}$.
Further, the mismatches of the SNRs and the noise signals might also lead to estimation errors,
and it seems that such mismatches affect the $\mathrm{DRR}$ estimate stronger than the $T_\mathrm{60}$ estimate.

Additionally, the proposed $(T_\mathrm{60}, \mathrm{DRR})$ estimator is
tested with the ACE multi-microphone data, but only one channel (here the first channel \emph{ch1})
is selected to perform the same estimation process.
The overall trend of the estimation results as shown in~Figure~\ref{fig:multi_ch1}
is similar to previous results, which serves as verification of our approach on a different (and larger) test set.
Again, the median values of $E_{T_\mathrm{60}}$ are near to 0~ms and the percentiles are within $\pm$250~ms,
and the median values of $E_{\mathrm{DRR}}$ are between -1~dB and -2~dB with $\pm$2.5~dB percentiles.
Consistent performances across noise types and SNRs indicate the importance of 
exploiting training data with a high amount of variability for a discriminative model
in order to achieve robustness in adverse conditions.

The computational cost of our approach is quantified in terms of the real-time factor~(RTF), 
defined as the ratio between the time taken to process a sentence and the length of the sentence.
Two components in our approach contribute most to the overall complexity, 
i.e., the calculation of Gabor features and the forward-run of the neural net (cf.~Figure~\ref{fig:rir_distru}). 
For optimization of the first component, the 2D convolution of spectrograms with Gabor filters 
was replaced by multiplication with a universal matrix.
Since the proposed MLP estimator operates on a GPU (cf.~Section~\ref{ssec:plat}),
the computational complexity is measured in frames per second~(FPS) with the frame length of 25~ms and overlapping of 10~ms.
A rough transfer from FPS to RTF can be computed by $\mathrm{RTF} =  1\, \mathrm{s} / ( \mathrm{FPS} \cdot 10\, \mathrm{ms} ) = 100 / \mathrm{FPS}$.
With an average GPU speed of 23736~FPS, an average RTF of 0.0042 is obtained.
In summary, the average RTF of the proposed estimator for the single-microphone scenario (4500 utterances)
is $0.0578+0.0042=0.0620$ (providing both $\widehat{T_\mathrm{60}}$ and $\widehat{\mathrm{DRR}}$),
while the RTFs of baseline $T_\mathrm{60}$ estimator~\cite{LYJV10} and $\mathrm{DRR}$ estimator~\cite{FC10} are 0.0483 and 0.3101, respectively.

\section{Conclusion}
\label{sec:conclu}
This contribution presented a novel method for $T_\mathrm{60}$ and $\mathrm{DRR}$ in a blind and joint way
using an MLP for classification.
It has been shown that the proposed method is capable of accurately estimating $T_\mathrm{60}$ and $\mathrm{DRR}$ 
in the context of the ACE Challenge using single-microphone, fullband speech signals.
The estimation errors of $T_\mathrm{60}$ and $\mathrm{DRR}$ cover a relatively small range of $\pm$250~ms and $\pm$2.5~dB
with corresponding median values of nearly 0~ms and -1.5~dB on average, respectively.
Furthermore, compared to the baseline approaches that only estimate either $T_\mathrm{60}$ or $\mathrm{DRR}$ estimation
at a time,
the computational complexity of the proposed estimator is significantly lower
since the signal processing for feature extraction and the forward-run of the neural net are not very demanding in terms of computational cost, 
and since the $T_\mathrm{60}$ and $\mathrm{DRR}$ are estimated simultaneously.
%


%

\end{sloppy}
\end{document}